\documentclass[11pt]{article}
\usepackage[sc]{mathpazo}
\usepackage{fullpage}
\usepackage[authoryear,sectionbib,sort]{natbib}
\usepackage[boxruled,linesnumbered]{algorithm2e}
\usepackage{amsmath}
\usepackage[utf8]{inputenc}
\usepackage{lineno}
\usepackage{titlesec}
\usepackage{graphicx}
\usepackage{subcaption}
\usepackage{booktabs}
\usepackage{multirow}
\usepackage{multicol}
\usepackage{here}
\usepackage{adjustbox}
\usepackage{mathtools}
\usepackage{url}

\titleformat{\section}[block]{\Large\bfseries\filcenter}{\thesection}{1em}{}
\titleformat{\subsection}[block]{\Large\itshape\filcenter}{\thesubsection}{1em}{}
\titleformat{\subsubsection}[block]{\large\itshape}{\thesubsubsection}{1em}{}
\titleformat{\paragraph}[runin]{\itshape}{\theparagraph}{1em}{}[. ]

\title{New confidence interval methods for Shannon index}

\author{Gabriel R. Palma$^{2}$ \and 
Silvio S. Zocchi$^{1}$ \and
Wesley A.C. Godoy$^{1}$ \and
Jorge A. Wiendl$^{1}$ }

\date{\today}
\DeclareUnicodeCharacter{2212}{-}
\begin{document}

\maketitle

\noindent{} 1. University of S\~{a}o Paulo, Piracicaba, Brazil;
\noindent{} 2. Maynooth University, Maynooth, Ireland;

\section*{Abstract}

\begin{enumerate}
    \item Several factors affect the structure of communities, including biological, physical and chemical phenomena, impacting the quantification of biodiversity, measured by diversity indexes such as Shannon's entropy. Then, once a point estimate is obtained, confidence intervals methods such as the bootstrap ones are often used. These methods, however, can have different performances, which many authors have revealed in the last decade. Furthermore, problems such as the asymmetry of the distribution of estimates and the possibility of Shannon's diversity index estimator bias can lead to incorrect recommendations to the research community. Thus, we propose two methods and compare them with seven others using their performances to face these problems.
    
    \item The first idea uses the credible interval (CI) method to build a bootstrap confidence interval. The second one starts by correcting the bias and then uses an asymptotic approach. We considered $27$ community structures representing scenarios with high dominance, high codominance or moderate dominance, the number of species equal to 4, 20 or 80 and 10, 50 or 500 individuals to compare their performances. Then, we generated 1000 samples, built $95\%$ confidence intervals, and calculated the percentage of times they included the community diversity index (coverage percentage) for each community structure.
    
    \item Our results showed the feasibility of both proposed methods to estimate Shannon's diversity. The simulation study revealed the bootstrap-t technique had the best performance, i.e., best coverage percentage, compared with the other methods. Finally, we illustrate the methodology by applying it to an original aphid and parasitoid species dataset. 
    
    \item We recommend the bootstrap-t when the community structure analysed is similar to the simulated ones. Also, the methods provided high performance for the high dominance scenarios.
    
\end{enumerate}
  
\textit{Keywords}: Diversity, statistical inference and bootstrap methods.

\section{Introduction}
The study of biodiversity involves interdisciplinary approaches aiming to understand the variety of life \citep{NOSS1990, Oconnor2020}. These studies resulted in several applications including biological control and pollination in agriculture to a clear notion of the ecosystem services that biodiversity promotes~\citep{Godfray1994, Williams1996, Altieri1999, Goodell2009, Mitchell2014}. Another aspect that highlights the importance of biodiversity is the comparison of impacts on natural communities, such as human impacts and others~\citep{Magurran2004, gotelli2001, Magurran2005, Magurran2010, mcglinn2019}. Researchers can also obtain measures of biodiversity in different scales depending on the study. For example, a local scale ($\alpha$ diversity) comprehends a community in a specific habitat~\citep{Magurran2004, gotelli2001}. Also, a regional scale ($\beta$ diversity) comprise the variety of species in different geographic areas. Finally, the landscape scale ($\gamma$ diversity) aim to obtain an overall species diversity within geographic areas~\citep{Levin2013, Magurran2004, Ricklefs2016}.

The most common indexes used in these different scales comprise Shannon's, Simpson's, and Chao's index~\citep{Magurran2004}, which appears in many studies as measures of biodiversity. Among these methods, to study local diversity ($\alpha$), Shannon's and Pielou's indices are the most popular ones to obtain information about the diversity and evenness of natural communities~\citep{Magurran2004}. A single index focus on some aspects of biodiversity. For example, Shannon's entropy and Pielou's index helps to understand some components of a community related to dominance, richness, diversity and evenness~\citep{Zar1999, Magurran2004, Levin2013}. However, simply computing these statistics based on a sample is not enough to  obtain accurate information in a location. 

For that, several methods aim to estimate such indices varying in scope from points to interval estimates. Among the literature of biodiversity, the usage of interval estimation is a common task including different methods, such as parametric or non-parametric techniques \citep{Davidson1997, Carvalho2019, Kamiyama2020, Willett2020, StLaurent2020, Guevara2020}. 
On the non-parametric approach, the bootstrap techniques are the most used for this task comprehending multiple confidence interval methods such as bootstrap t, bias corrected-accelerated, bias-corrected percentile, percentile, standard percentile and empirical corrected methods \citep{Fritsch2000, Pla2004, McDonald2010, Scherer2013, Pesenti2016}. Previous studies evaluated the performance of the bootstrap confidence interval methods to suggest a recommendation, mainly to estimate Shannon's diversity, for biologists that aim the use a quantitative approach to study biodiversity \citep{Pla2004, Pesenti2016}.  

To evaluate the performance of confidence interval methods these studies performed simulations using different community structures. These structures have distinct species abundance distributions, number of species, and individuals in a certain location. Thus, they consider the index calculated on these communities as the parameters target of estimation. Moreover, by taking samples from them, it is possible to obtain statistics, such as width and coverage percentage showing the performance of each confidence interval method. The first statistic is the difference between upper and lower interval limits and the other is the number of times that the interval contains the parameter divided by the total number of samples. However, the restricted number of community structures simulated, can affect the method recommendations because, as \cite{Hutcheson1970}, \cite{shenton1969}, \cite{Pla2004} and \cite{Zhang2012} mentioned, the species abundance distribution affect the bias of Shannon's index. 

Among these studies, there are several results suggesting gaps in the literature that requires attention. These gaps include the creation of simulation studies with communities biologically meaningful. Also, more studies focused on the bias of Shannon's index \citep{shenton1969, Hutcheson1970, Pla2004, Zhang2012}, and alternative methods that deal with the fact that bootstrap confidence interval methods are inadequate for situations where there is asymmetry and multimodality in the distribution of estimates. 

Therefore, we propose two new confidence interval methods for Shannon's index. The basis of the first one is a bootstrap approach, and the second is an asymptotic approach. They deal with, respectively, asymmetry (i.e. a distribution skewed for the right or left) and Shannon's index bias (i.e. Point estimation error of the parameter $H$). Also, we aim to compare it with bootstrap t (t), bias corrected-accelerated (BCA), bias-corrected percentile (BC), percentile (Perc), standard percentile (SPerc) and empirical corrected (EC) methods using a simulation study with biological meaningful community structures. Given that high dominance, moderate dominance and codominance patterns are commonly encountered in community structures restricted to a local scale, the simulated scenarios were based on a degree of dominance according to the codominance and evenness indices \citep{Pielou1966, Gray2021}. Finally, to support our recommendation, we presented a case study of an aphid monitoring system emphasising the scenarios presented in the simulation study.

\section{Methods}
In this section we describe the proposed confidence interval methods and compare them with some traditional bootstrap methods. 
Let 
\begin{equation}
\mathbf{x} =\left(  \begin{array}{cccc}
Sp_1&Sp_2&\ldots&Sp_{k} \\
n_1& n_2& \ldots&n_{k} \\
\end{array}
\right)
\label{sample}
\end{equation} be the observed species frequency distribution in a sample, where: $k$ is the number of species; $n_j$ $\left( j = 1, \ldots, k\right)$ is the frequency of individuals of species $Sp_j$ and $n ={\sum_{j=1}^{k}n_j}$ is the total number of individuals. 
 Consider  that $H$ is the parameter of interest and  $\widehat{H} = -\sum^{k}_{j = 1}\frac{n_j}{n}\log\frac{n_j}{n}$ ($n_j \neq 0$) its estimator. Then, as an example, the 7 individuals sample \begin{equation}
    \mathbf{x} =\left\{ Sp_1, Sp_2, Sp_2, Sp_4, Sp_3, Sp_2, Sp_3\right\}=\left(  \begin{array}{cccc}
    Sp_1 & Sp_2& Sp_3 & Sp_4 \\
    1    &    3&    2 & 1\\
    \end{array} 
\right)
    \label{sample_example}
\end{equation}
will give the estimate $\widehat{H}=-\left(\frac{1}{7}\log \frac{1}{7}+\frac{3}{7}\log \frac{3}{7}+\frac{2}{7}\log \frac{2}{7}+\frac{1}{7}\log \frac{1}{7}\right)= 1.2770$. 

Despite its frequent use, the estimator $\widehat{H}$ generally underestimates $H$ \citep{Pla2004}, but, taking into account  the  2-nd order series expansion expressions for the expected bias of $\widehat{H}$  and variance of $\widehat{H}$, $Var(\widehat{H})$,  presented by \cite{Hutcheson1970}, \cite{shenton1969}, \cite{Pla2004} and \cite{Zhang2012}, an approximately unbiased estimator of $H$ will be
\begin{equation}
    \widehat{H}^{\prime} = 
    \widehat{H} + 
    \frac{k - 1}{2n} - \frac{1}{12n^2} \left( 1 - \displaystyle\sum^{k}_{j = 1}\frac{n}{n_j}\right),
\end{equation} 
with approximate variance  
\[
\widehat{Var}(\widehat{H}^{\prime})  \approx \frac{1}{n}\left(\displaystyle\sum^{k}_{j=1}\frac{n_{j}}{n}\left(\log{\frac{n_{j}}{n}}\right)^{2} - \widehat{H}^2\right) + \frac{k - 1}{2n^2}.
\]
Furthermore, according to the authors $\widehat{H}$ is asymptotically normal distributed and thus, so will be $\widehat{H}^{\prime}$. 
Then, the proposed   ($1-\alpha$)100\% {\bf asymptotically corrected} confidence interval for $H$ will  be
\begin{equation}
    CI_{(1-\alpha)100\%}(H) = \widehat{H}^{\prime} \pm \sqrt{\widehat{Var}(\widehat{H}^{\prime})}~
t_{\left\{1-\frac{\alpha}{2}, n - 1\right\}},
    \label{CI_asymp}
\end{equation} 
where $t_{\left\{1-\frac{\alpha}{2}, n - 1\right\}}$ is the ($1-\frac{\alpha}{2}$)-th quantile of the $t$ distribution with $n - 1$ degrees of freedom. Now, considering the dataset~(\ref{sample_example}), we will have $\widehat{H}^{\prime}=1.5233$, $\widehat{Var}(\widehat{H}^{\prime}) = 0.06020$, $t_{\left\{0.975,6\right\}}=2.4469$ and $CI_{95\%}(H)=(0.9230, 2.1237)$, noticing that for small samples like in this case ($n=7$), the expression (\ref{CI_asymp}) gives only an approximate interval. 

Hereinafter, we describe the fundamental idea of bootstrap  methodology and how to use it to build a percentile confidence interval for $H$. The first step is to extract $B$ random samples with replacement of (\ref{sample}),  known as bootstrap samples \citep{EfroTibs93,Davidson1997}. Let the $b$-th bootstrap sample $(b = 1, \dots, B )$ be
\begin{equation}
    \mathbf{x}^{(b)} =\left(  \begin{array}{cccc}
    Sp_1&Sp_2&\ldots&Sp_{k} \\
    n_1^{(b)}& n_2^{(b)}& \ldots&n_{k}^{(b)} \\
    \end{array}
\right), \nonumber
    \label{bootstrap_sample}
\end{equation} 
where $n^{(b)}_j$ is the frequency of species $Sp_j$ ($j = 1, \dots, {k} $) in the sample. The following step is to compute the statistic of interest, $\widehat{H}$, in our case, for each bootstrap sample, named $\widehat{H}^{(b)}$, forming the set 
\begin{equation}
 (\widehat{H}^{(1)}, \widehat{H}^{(2)},  \ldots, \widehat{H}^{(B)}).   
 \label{bootstrap_set}
\end{equation}
As an example, the $5$-th bootstrap sample extracted from~(\ref{sample_example}) could be
\begin{equation}
\mathbf{x}^{(5)} =
\left\{ Sp_3, Sp_2, Sp_2, Sp_3, Sp_2, Sp_2, Sp_3\right\} = \left(  \begin{array}{cccc}
Sp_1 & Sp_2 & Sp_3 & Sp_4 \\
 0   &  4   &  3   &  0   \\
\label{shannon_example}
\end{array}
\right),
\\
  \end{equation}
 leading to the estimate  $\widehat{H}^{(5)}= -\left(\frac{4}{6}\log\frac{4}{6}+\frac{3}{6}\log\frac{3}{6}\right)=0.6829$.
\nonumber
  \label{asdas}\nonumber
Now, given the set of bootstrap estimates (\ref{bootstrap_set}), the \textbf{percentile bootstrap}  $100(1-\alpha) \%$ confidence interval  for $H$  is $CI(H)_{100(1-\alpha)\%}=
\left(P_{\frac{\alpha}{2}},
 P_{1-\frac{\alpha}{2}}\right)$, where $P_{\frac{\alpha}{2}}$ and
 $P_{1-\frac{\alpha}{2}}$ are the  $\frac{\alpha}{2}$-th and $(1-\frac{\alpha}{2})$-th percentiles of (\ref{bootstrap_set}), respectively. 
 
However, despite the popularity of this and other bootstrap methods in building confidence intervals, when the distribution of bootstrap estimates is skewed, the use of credible intervals shall be more appropriate.  Essentially, a one mode $(1-\alpha)100\%$ \textbf{credible interval} (CrI) is the interval with the smallest width such that the percentage of values of (\ref{bootstrap_set}) pertaining to it is greater than or equal to $(1-\alpha)100\%$.

From this point on, we describe the simulation methodology to compare the performance of the proposed methods, credible interval (CrI) and asymptotically corrected (AC), with the methods:  bootstrap t (t); bias corrected-accelerated (BCA); bias-corrected percentile (BC); percentile (Perc); standard percentile (SPerc) and empirical corrected (EC). 

The first step is to set the \textbf{number of species} in the community, $K$, considered here, equal to $4$, $20$, or $80$. Low numbers of species $(K<30)$ are common in community structures of agricultural landscapes, which exhibits low number of pests in response to poor diversity of host plants \citep{guo2019}. In larger areas with an heterogeneity old-growth rainforest, though, a larger number of species, e.g. $K=80$, are more likely to occur   \citep{vanclay1989, ojo1996, Mulder2004, Wubs2018}. 

Given $K$, the following step is to set the relative abundance of species, $Sp_j$  ($j = 1, 2, \ldots, K$), in these community structures, given by
\[p_j=\frac{1}{100K}+\frac{99}{100}\frac{j^{-\nu}}{\sum_{j=1}^K j^{-\nu}}~~~~(j=1,\ldots,K),\]
where $\nu\ge0$ is a parameter that is related to the level of dominance and codominance in the community (Table~\ref{simulationScenarios}). Note that $p_1 > p_2 > \cdots > p_{K - 1} > p_{K} > \frac{1}{100K}$, $\sum_{j=1}^K p_j=1$ and $\frac{1}{100K}$ is the minimum detection probability of any species. The choice of $\nu$ was set for \textbf{three different scenarios}: 1. high dominance; 2. high codominance and 3. moderate dominance. Given $K$, the values of $\nu$ for the first and third scenarios were the ones that the Pielou´s evenness index, $J = \frac{H}{\log{K}}$, where 0.15 and 0.9, respectively. On the other hand, given $K$, the chosen $\nu$ value for the second scenario was the one that maximized the codominance index by \cite{Gray2021},
\[
\label{codominance.degree}
C_{k_{c}}=\frac{k_{c}}{\displaystyle\sum_{j=1}^{k_{c}}\frac{1}{p_{j}}} - p_{k_{c}+1}
\]
where $2\le k_c<K$
is the number of codominance species, considered here equal to 2, that is
\[
C_{k_c}=C_2=\frac{2}{\frac{1}{p_1}+\frac{1}{p_2}}-p_3.
\]
$C_2$ is, then, the difference between the harmonic mean of $p_1$ and $p_2$, and $p_3$.

\begin{table}[ht]
	\caption{Chosen values of $\nu$ and values of Pielou's evenness index, $J$, Shannon's diversity index, $H$ and codominance index, $C_2$, for three different scenarios: 1. high dominance; 2. high codominance and 3. moderate dominance, given the number of species in the community, $K$.}
	\label{simulationScenarios}
	\centering
	\begin{adjustbox}{max width=\textwidth}
	\begin{tabular}{lccccc}
		\hline
		& & \multicolumn{4}{c}{Parameters}
		\\\cmidrule{2-6}
		Scenarios&
	    \multicolumn{1}{c}{$K$} & 
	    \multicolumn{1}{c}{$\nu$} & 
		\multicolumn{1}{c}{$J$} &
		\multicolumn{1}{c}{$H$} &
		\multicolumn{1}{c}{$C_2$}
		\\
		\hline
		& 4& 4.9618& 0.15& 0.21& 0.0574\\
		1. High dominance& 20& 3.7901& 0.15&0.45& 0.1079\\
		& 80& 3.2263 & 0.15& 0.66& 0.1399\\[0.2cm]
		
		& 4& 1.8480& 0.69& 0.96 & 0.2029\\
		2. High codominance & 20& 2.1700& 0.42& 1.27&0.1811\\
		& 80& 2.2250& 0.30& 1.34 & 0.1788\\[0.2cm]
		
		& 4& 0.9913& 0.90& 1.25& 0.1577\\
		3. Moderate dominance& 20& 0.8210& 0.90& 2.70& 0.0687\\
		& 80& 0.7612& 0.90& 3.94& 0.0367\\[0.2cm]
		\hline\\
	\end{tabular}
	\end{adjustbox}
\end{table}

For each community structure, i.e., combination of scenario and value of $K$, presented on Table~\ref{simulationScenarios}, we extracted $N=1000$ random samples  with \textbf{size} $n =10$, 50 or 500 individuals and for each generated sample,  $95\%$ confidence intervals for $H$ were built based on the described  methods. Note that for bootstrap methods, we used $B=1000$ bootstrap samples. Also, we assumed that missing species in samples of community structures and bootstrap samples are not considered for computing Shannon's index. The equation~\ref{shannon_example} shows an example of it.

Then, in order to compare their performances, for each set of $N=1000$ confidence intervals, we computed the \textbf{coverage percentage}, i.e., the percentage of times that $H$ belongs to the confidence interval and the \textbf{average width}, i.e., the mean of the widths of the 1000 confidence intervals. It is expected that a good method will produce a coverage percentage close to 95\% and a narrow average width. 
\section{Results}

\begin{table}[ht]
	\caption{ Coverage percentages and average widths (between parenthesis) for different 95\% confidence interval methods for $H$,  considering: scenarios (Scen.) with 1. high dominance, 2. high codominance or 3. moderate dominance;  different sample sizes, $n=10$, 50 or 500 and different community number of species, $K=4$, 20 or 80. The results were based on 1000 samples, 1000 bootstrap samples (for bootstrap methods) and 25 subsamples of bootstrap samples for the $t$ method. Note: $\bar{k}$ is the average number of species in the samples. }
	\label{ShannonTable}
	\centering
	\begin{adjustbox}{max width=\textwidth}
	\begin{tabular}{crrrrrrrrrrr}
		\hline
		& & \multicolumn{10}{c}{Bootstrap method}
		
		\\\cmidrule{5-10}
		Scen.&
		$n$  & 
		\multicolumn{1}{c}{$K$} & 
		\multicolumn{1}{c}{$\bar{k}$}&
		\multicolumn{1}{c}{Perc} & 
		\multicolumn{1}{c}{BCA} & 
		\multicolumn{1}{c}{t} &
		\multicolumn{1}{c}{BC}&
		\multicolumn{1}{c}{SPerc}&
		\multicolumn{1}{c}{EC}&
		\multicolumn{1}{c}{Crl}&
		\multicolumn{1}{c}{AC}
		\\
		\hline
    &  & 4 & 1.40 & 36 (0.24) & \textbf{98} (0.65) & \textbf{99} (1.08) & 36 (0.20) & 36 (0.31) & 30 (0.31) & 36 (0.23) & 36 (0.37) \\[-0.3cm]
    & 10 & 20 & 1.78 & 64 (0.45) & \textbf{99} (0.69) & \textbf{99} (0.98) & 64 (0.40) & 64 (0.55) & 60 (0.55) & 64 (0.43) & 63 (0.65) \\[-0.3cm]
   &  & 80 & 2.15 & 44 (0.60) & 38 (0.67) & \textbf{96} (1.27) & 43 (0.54) & 78 (0.70) & 70 (0.70) & 44 (0.56) & 75 (0.82) \\[-0.1cm]

  & & 4 & 2.25 & \textbf{87} (0.31) & \textbf{99} (0.37) & \textbf{99} (0.60) & 65 (0.29) & \textbf{87} (0.33) & \textbf{85} (0.33) & \textbf{87} (0.29) & \textbf{87} (0.34) \\[-0.3cm]
  1 & 50 & 20 & 3.42 & \textbf{80} (0.47) & \textbf{83} (0.49) & \textbf{99} (0.68) & 78 (0.47) & \textbf{84} (0.48) & \textbf{84} (0.48) & 77 (0.45) & \textbf{84} (0.49) \\[-0.3cm]
  & & 80 & 4.21 & 75 (0.53) & \textbf{81} (0.52) & \textbf{95} (0.70) & 78 (0.52) & 76 (0.54) & \textbf{84} (0.53) & 74 (0.51) & \textbf{85} (0.55) \\  [-0.1cm]
  
  & & 4 & 3.81 & \textbf{93} (0.13) & \textbf{96} (0.14) & \textbf{98} (0.15) & \textbf{94} (0.13) & \textbf{93} (0.14) & \textbf{95} (0.14) & \textbf{92} (0.13) & \textbf{94} (0.14) \\[-0.3cm] 
  & 500 & 20 & 8.90 & \textbf{91} (0.19) & \textbf{93} (0.19) & \textbf{96} (0.21) & \textbf{93} (0.19) & \textbf{91} (0.19) & \textbf{93} (0.19) & \textbf{90} (0.18) & \textbf{93} (0.19) \\[-0.3cm] 
  & & 80 & 12.45 & \textbf{83} (0.21) & \textbf{87} (0.21) & \textbf{94} (0.23) & \textbf{89} (0.21) & \textbf{83} (0.21) & \textbf{94} (0.21) & \textbf{82} (0.21) & \textbf{90} (0.21) \\[0.2cm]

  & & 4 & 2.92 & 62 (0.78) & 72 (0.68) & \textbf{86} (1.10) & 71 (0.68) & 72 (0.84) & 77 (0.84) & 60 (0.68) & \textbf{85} (0.97) \\[-0.3cm] 
   & 10 & 20 & 3.40 & 43 (0.88) & 44 (0.71) & \textbf{94} (1.35) & 43 (0.71) & 45 (0.93) & 73 (0.93) & 44 (0.77) & 79 (1.09) \\[-0.3cm] 
  & & 80 & 3.43 & 17 (0.88) & 42 (0.71) & \textbf{96} (1.37) & 37 (0.71) & 42 (0.94) & 75 (0.94) & 25 (0.79) & \textbf{80} (1.11) \\[-0.1cm]
  
  & & 4 & 3.92 & \textbf{90} (0.45) & \textbf{90} (0.44) & \textbf{95} (0.53) & \textbf{90} (0.43) & \textbf{91} (0.46) & \textbf{89} (0.46) & \textbf{90} (0.44) & \textbf{91} (0.47) \\[-0.3cm] 
  2 & 50 & 20 & 7.35 & 70 (0.63) & \textbf{84} (0.62) & \textbf{93} (0.78) & \textbf{80} (0.59) & 72 (0.63) & \textbf{86} (0.63) & 72 (0.62) & \textbf{85} (0.65) \\[-0.3cm] 
  & & 80 & 7.71 & 58 (0.64) & \textbf{80} (0.63) & \textbf{90} (0.80) & 76 (0.61) & 60 (0.65) & \textbf{84} (0.65) & 58 (0.64) & \textbf{82} (0.67) \\ [-0.1cm]
  
  & & 4 & 4.00 & \textbf{95} (0.15) & \textbf{96} (0.15) & \textbf{96} (0.15) & \textbf{95} (0.14) & \textbf{95} (0.15) & \textbf{95} (0.15) & \textbf{95} (0.14) & \textbf{95} (0.15) \\[-0.3cm] 
  & 500 & 20 & 16.74 & \textbf{89} (0.24) & \textbf{94} (0.24) & \textbf{96} (0.26) & \textbf{94} (0.24) & \textbf{90} (0.24) & \textbf{93} (0.24) & \textbf{88} (0.24) & \textbf{94} (0.24) \\[-0.3cm] 
  & & 80 & 23.80 & 73 (0.26) & \textbf{89} (0.26) & \textbf{92} (0.29) & \textbf{87} (0.26) & 74 (0.26) & \textbf{92} (0.26) & 73 (0.26) & \textbf{89} (0.26) \\[0.2cm]

  & & 4 & 3.50 & 54 (0.72) & 49 (0.47) & \textbf{88} (1.04) & 49 (0.49) & 56 (0.77) & \textbf{85} (0.76) & 54 (0.62) & \textbf{96} (0.88) \\[-0.3cm] 
  & 10 & 20 & 7.07 & 0 (0.80) & 0 (0.14) & 72 (0.96) & 0 (0.14) & 0 (0.82) & 69 (0.82) & 0 (0.74) & 60 (1.02) \\[-0.3cm] 
  & & 80 & 8.74 & 0 (0.74) & 0 (0.03) & 0 (0.84) & 0 (0.02) & 0 (0.77) & 0 (0.77) & 0 (0.69) & 0 (1.00) \\ [-0.1cm]

  & & 4 & 4.00 & \textbf{91} (0.31) & \textbf{92} (0.27) & \textbf{93} (0.33) & \textbf{91} (0.26) & \textbf{94} (0.31) & \textbf{89} (0.31) & \textbf{92} (0.29) & \textbf{93} (0.32) \\[-0.3cm]
  3 & 50 & 20 & 16.28 & 4 (0.47) & 23 (0.20) & \textbf{91} (0.51) & 20 (0.20) & 7 (0.47) & 78 (0.47) & 6 (0.46) & \textbf{89} (0.49) \\[-0.3cm]
  & & 80 & 30.53 & 0 (0.45) & 0 (0.01) & 26 (0.50) & 0 (0.00) & 0 (0.46) & 70 (0.46) & 0 (0.45) & 17 (0.49) \\[-0.1cm]

  & & 4 & 4.00 & \textbf{94} (0.09) & \textbf{95} (0.09) & \textbf{96} (0.10) & \textbf{95} (0.09) & \textbf{95} (0.09) & \textbf{94} (0.09) & \textbf{94} (0.09) & \textbf{96} (0.09) \\[-0.3cm]
  & 500 & 20 & 20.00 & \textbf{83} (0.14) & \textbf{94} (0.14) & \textbf{96} (0.15) & \textbf{94} (0.13) & \textbf{86} (0.15) & \textbf{90} (0.15) & \textbf{84} (0.14) & \textbf{95} (0.15) \\[-0.3cm] 
  & & 80& 77.05 & 3 (0.19) & 31 (0.07) & \textbf{95} (0.20) & 32 (0.07) & 4 (0.19) & 57 (0.19) & 4 (0.18) & \textbf{95} (0.19) \\[0.2cm]  
   \hline

	\end{tabular}
	\end{adjustbox}
	
\end{table}

The simulation results are presented in Table~\ref{ShannonTable}. We expected that the coverage percentages would be close to 95$\%$ and considered here, as acceptable, values greater or equal to 80$\%$.
Following this consideration, the bootstrap t, AC, bootstrap BCA, bootstrap EC, bootstrap Perc, bootstrap SPerc and Crl methods have acceptable coverage percentages in 89, 74, 63, 59, 41, 41 and 37$\%$ of the 27 considered community structures, respectively. Based on these results, we recommend the bootstrap t. 

Now, considering communities with a high dominance scenario, though, the best two methods were the bootstrap t and BCA (100 and 89$\%$ of the 9 simulated community structures with acceptable coverage percentages, respectively). 
Moreover, for the communities with a high codominance scenario, the best two were bootstrap t and AC (100 and 89$\%$ of the 9 simulated community structures with acceptable coverage percentages, respectively). 
Finally, for the moderate dominance scenario, the bootstrap t and the AC (both with 67$\%$ of the 9 simulated community structures with acceptable coverage percentages) were the best two methods. 

\begin{table}[ht]
	\caption{Abundances of parasitoids and wheat crop aphids in three periods in a monitoring programme at Area II ($710~m$ altitude, $28^\circ~11^{\prime}~42.8^{\prime\prime}$ S and $52^\circ~19^{\prime}~30.6^{\prime\prime} W$) of the Embrapa Trigo experimental station, located in Coxilha, RS, Brazil. 
	}
	\label{datasetTable}
	\centering
	\begin{adjustbox}{max width=\textwidth}
	\begin{tabular}{lccc}
		\hline
		Species &\multicolumn{1}{c}{20-26/08/14} &	\multicolumn{1}{c}{22-28/01/16}  & \multicolumn{1}{c}{02-08/04/18}\\	
\hline 
\\[-0.5cm]				
Parasitoids \\				
\textit{~~~~Aphidius rhopalosiphi} (DeStefani, 1902) &	2 &	0 &	 0 \\	
\textit{~~~~Aphidius platensis} (Brãthes, 1913) &	1 &	0 &	 3\\	
\textit{~~~~Diaeretiella rapae} (Mc'Intosh, 1855) &	1 &	0 &	 1\\	
Aphids\\				
\textit{~~~~Metapolophium dirhodum} (Walker, 1849) & 0 & 1 &  0\\				
\textit{~~~~Rhopalosiphum padi} (Linnaeus, 1758) &  6 &  387~~~~ &  0\\ 				
\textit{~~~~Rhopalosiphum maidis} (Luzhetzki, 1960) & 0 & 0 &  1\\ 				
\textit{~~~~Rhopalosiphum rufiabdominalis} (Sasaki, 1899) & 0 & 1 &  0\\ 				
		
\textit{~~~~Sitibion avenae} (Fabricius, 1775) & 0 & 10 &  1\\				
\hline &  &  & 				
\\[-0.6cm] 			
$~~~~n$ &  10~~ &  399~~~~ &  6~~\\				
$~~~~\widehat{H}$ &  ~~~~1.09 &  ~~~~0.15 &  ~~~~1.24\\
$~~~~\widehat{J}=\displaystyle\frac{\widehat{H}}{\log k}$ &  ~~~~0.79 &  ~~~~0.11 &  ~~~~0.90\\		
				
$~~~~\widehat{C}_2$ &	 ~~~~~0.20 &	 ~~~~~0.046 &	 ~~~~~0.083\\[0.2cm]	
\hline
	\end{tabular}
	\end{adjustbox}
	
\end{table}

As an example, we will now calculate the confidence intervals of $H$ for a community of parasitoids and wheat crop aphids presented in Table~\ref{datasetTable}.
As the statistics, $k$, $\widehat{H}$, $\widehat{J}$ and $\widehat{C}_2$, calculated for the samples collected during the three periods were similar to the ones presented in the simulation study, the bootstrap t method was selected.
The first period shows a high codominance of species \textit{Rhopalosiphum padi} and \textit{Aphidius rhopalosiphi} and $CI_{95\%}(H)=(0.73, 2.32)$.
The second, a high dominance of \textit{Rhopalosiphum padi} and $CI_{95\%}(H)=(0.09, 0.25)$. The last, a moderate dominance and $CI_{95\%}(H)=(1.12, 3.12)$. 

\section{Discussion}
We proposed the credible interval (CrI) and an asymptotically corrected (AC) method to estimate confidence intervals for Shannon's index. Also, we compared them with commonly used methods for interval estimation of this diversity index. Therefore, the recommended method was the bootstrap-t. 

The example presented showed how to use the results obtained from the simulated scenarios in real-world data. These scenarios reflect real conditions of aphids exploring wheat culture \citep{muller1999, von2011} because our simulation uses the degree of dominance as conditions to generate community structures. It is emphasised by the fact that environments with lower resource heterogeneity tend to present community structure with high dominance or codominance \citep{wilsey2002, Tsakalakis2020, Chaves2021}. These patterns are not restricted to the presented taxa, and they could be observed in other insect orders, such as Fruit Flyes and Dung insect communities \citep{hanski2014, Araujo2021}.

Now, by analysing papers similar to ours, we found that \cite{Pesenti2016} have used the average width and computational speed as a criterion of method selection. They concluded that the bootstrap percentile must be recommended based on their simulation study. Table~\ref{ShannonTable} shows that if we use their criteria and compare the bootstrap percentile, bias-corrected, bias-corrected accelerated and t, we agree with their conclusions. However, by introducing the other methods in the comparison, the proposed asymptotic corrected method has better performance because it is $4053$ times faster than the bootstrap percentile. 

Moreover, wide or fewer intervals are not necessarily the best ones as they may not contain the community index. Is more important than the produced intervals contain most of the times this index, which is evaluated through the percentage of coverage. Given that said, table~\ref{ShannonTable} shows a clear example of it considering the second scenario with $n = 10$. The methods with higher average width contain Shannon's index in the community more times than the ones with wide intervals.


Finally, our findings also suggest that the community structure can influence the selection of methods for the construction of confidence intervals for Shannon's index. Our main contribution is an innovative approach to provide recommendations for naturalists studying local diversity. Also, our simulation study helps the selection of appropriate confidence interval methods to estimate Shannon's index, and this approach can be extended to any diversity index. However, the limitations of our work rely on the restricted simulated scenarios leading us to produce limited recommendations for community structures. So, in future work, we will explore more biological scenarios.

\section{Conclusion}
The simulation study showed that the bootstrap-t method produced a superior performance for $88\%$ of the simulated dataset for Shannon's diversity, $H$, indicating its superior overall performance as compared to the others and, thus, we recommend it when the community structure analysed is similar to the ones that we simulated. Finally, the methods provided more accurate estimations of Shannon's diversity for the high dominance scenario.
\bibliographystyle{chicago}
\bibliography{ref.bib}

\begin{thebibliography}{}

\bibitem[\protect\citeauthoryear{Altieri}{Altieri}{1999}]{Altieri1999}
Altieri, M.~A. (1999).
\newblock The ecological role of biodiversity in agroecosystems.
\newblock {\em Agriculture, Ecosystems \& Environment\/}~{\em 74\/}(1), 19--31.

\bibitem[\protect\citeauthoryear{Carvalho, Santana, and Sampaio}{Carvalho
  et~al.}{2019}]{Carvalho2019}
Carvalho, F., D.~Santana, and M.~Sampaio (2019, 11).
\newblock Modeling overdispersion, autocorrelation, and zero-inflated count
  data via generalized additive models and bayesian statistics in an aphid
  population study.
\newblock {\em Neotropical Entomology\/}~{\em 49}, 40--51.

\bibitem[\protect\citeauthoryear{Chaves and Smith}{Chaves and
  Smith}{2021}]{Chaves2021}
Chaves, F.~A. and M.~D. Smith ({2021}, {OCT}).
\newblock {Resources do not limit compensatory response of a tallgrass prairie
  plant community to the loss of a dominant species}.
\newblock {\em {JOURNAL OF ECOLOGY}\/}~{\em {109}\/}({10}), {3617--3633}.

\bibitem[\protect\citeauthoryear{Davison and Hinkley}{Davison and
  Hinkley}{1997}]{Davidson1997}
Davison, A.~C. and D.~V. Hinkley (1997).
\newblock {\em Bootstrap Methods and Their Application}.
\newblock Cambrige, UK: Cambridge University Press.

\bibitem[\protect\citeauthoryear{de~Araujo, Martins, Fornazier, Uramoto,
  Ferreira, Zucchi, and Godoy}{de~Araujo et~al.}{2021}]{Araujo2021}
de~Araujo, M.~R., D.~d.~S. Martins, M.~J. Fornazier, K.~Uramoto, P.~S.
  Ferreira, R.~A. Zucchi, and W.~A. Godoy (2021).
\newblock Long-term fruit fly monitoring and impact of the systems approach on
  richness and abundance.
\newblock {\em The Canadian Entomologist\/}~{\em 153\/}(6), 682–701.

\bibitem[\protect\citeauthoryear{Efron and Tibshirani}{Efron and
  Tibshirani}{1993}]{EfroTibs93}
Efron, B. and R.~J. Tibshirani (1993).
\newblock {\em An Introduction to the Bootstrap}.
\newblock Number~57 in Monographs on Statistics and Applied Probability. Boca
  Raton, Florida, USA: Chapman Hall \& CRC.

\bibitem[\protect\citeauthoryear{Fritsch and Hsu}{Fritsch and
  Hsu}{2000}]{Fritsch2000}
Fritsch, K. and J.~Hsu (2000, 01).
\newblock Multiple comparison of entropies with application to dinosaur
  biodiversity.
\newblock {\em Biometrics\/}~{\em 55}, 1300--1305.

\bibitem[\protect\citeauthoryear{Godfray}{Godfray}{1994}]{Godfray1994}
Godfray, C. (1994, 12).
\newblock {\em Parasitoids: Behavioral and Evolutionary Ecology}.
\newblock William Street, Princeton, New Jersey, US: Princeton University
  Press.

\bibitem[\protect\citeauthoryear{Goodell}{Goodell}{2009}]{Goodell2009}
Goodell, P. (2009, 12).
\newblock Fifty years of the integrated control concept: The role of landscape
  ecology in ipm in san joaquin valley cotton.
\newblock {\em Pest management science\/}~{\em 65}, 1293--1297.

\bibitem[\protect\citeauthoryear{Gotelli and Colwell}{Gotelli and
  Colwell}{2001}]{gotelli2001}
Gotelli, N.~J. and R.~K. Colwell (2001).
\newblock Quantifying biodiversity: procedures and pitfalls in the measurement
  and comparison of species richness.
\newblock {\em Ecology letters\/}~{\em 4\/}(4), 379--391.

\bibitem[\protect\citeauthoryear{Gray, Komatsu, and Smith}{Gray
  et~al.}{2021}]{Gray2021}
Gray, J.~E., K.~J. Komatsu, and M.~D. Smith (2021).
\newblock Defining codominance in plant communities.
\newblock {\em New Phytologist\/}~{\em 230}, 1--15.

\bibitem[\protect\citeauthoryear{Guo, Fei, Potter, Liebhold, and Wen}{Guo
  et~al.}{2019}]{guo2019}
Guo, Q., S.~Fei, K.~M. Potter, A.~M. Liebhold, and J.~Wen (2019).
\newblock Tree diversity regulates forest pest invasion.
\newblock {\em PNAS\/}~{\em 116\/}(15), 7382--7386.

\bibitem[\protect\citeauthoryear{Hanski}{Hanski}{2014}]{hanski2014}
Hanski, I. (2014).
\newblock The dung insect community.
\newblock In {\em Dung beetle ecology}, pp.\  5--21. Princeton University
  Press.

\bibitem[\protect\citeauthoryear{Hutcheson}{Hutcheson}{1970}]{Hutcheson1970}
Hutcheson, K. (1970).
\newblock A test for comparing diversities based on the shannon formula.
\newblock {\em Journal of Theoretical Biology\/}~{\em 29}, 151--154.

\bibitem[\protect\citeauthoryear{Kamiyama, Bradford, Groves, and
  Guedot}{Kamiyama et~al.}{2020}]{Kamiyama2020}
Kamiyama, M., B.~Bradford, R.~Groves, and C.~Guedot (2020, 04).
\newblock Degree day models to forecast the seasonal phenology of
  \textit{Drosophila suzukii} in tart cherry orchards in the midwest us.
\newblock {\em PLOS ONE\/}~{\em 15}, 1--13.

\bibitem[\protect\citeauthoryear{Levin}{Levin}{2013}]{Levin2013}
Levin, S.~A. (2013).
\newblock {\em Encyclopedia of Biodiversity\/} (2$^{nd}$ ed.).
\newblock London, UK: Academic Press.

\bibitem[\protect\citeauthoryear{Magurran}{Magurran}{2004}]{Magurran2004}
Magurran, A.~E. (2004).
\newblock {\em Measuring Biological Diversity}.
\newblock Main Street, Malden, USA: Blackwell publishing.

\bibitem[\protect\citeauthoryear{Magurran}{Magurran}{2005}]{Magurran2005}
Magurran, A.~E. (2005).
\newblock Biological diversity.
\newblock {\em Current biology\/}~{\em 15\/}(4), 116--119.

\bibitem[\protect\citeauthoryear{Magurran}{Magurran}{2010}]{Magurran2010}
Magurran, A.~E. (2010).
\newblock Open access q \& a : What is biodiversity?
\newblock {\em BMC Biology\/}~{\em 8\/}(145), 1--4.

\bibitem[\protect\citeauthoryear{McDonald, Smith, Scot, and Dick}{McDonald
  et~al.}{2010}]{McDonald2010}
McDonald, C., R.~Smith, M.~Scot, and J.~Dick (2010, 06).
\newblock Using indices to measure biodiversity change through time.
\newblock In {\em 5th METMA Santiago de Compostela, Spain, June-July 2010},
  pp.\  1--5. International Workshop on Spatial-Temporal Modeling.

\bibitem[\protect\citeauthoryear{McGlinn, Xiao, May, Gotelli, Engel, Blowes,
  Knight, Purschke, Chase, and McGill}{McGlinn et~al.}{2019}]{mcglinn2019}
McGlinn, D.~J., X.~Xiao, F.~May, N.~J. Gotelli, T.~Engel, S.~A. Blowes, T.~M.
  Knight, O.~Purschke, J.~M. Chase, and B.~J. McGill (2019).
\newblock Measurement of biodiversity (mob): A method to separate the
  scale-dependent effects of species abundance distribution, density, and
  aggregation on diversity change.
\newblock {\em Methods in Ecology and Evolution\/}~{\em 10\/}(2), 258--269.

\bibitem[\protect\citeauthoryear{Mitchell and Onstad}{Mitchell and
  Onstad}{2014}]{Mitchell2014}
Mitchell, P. and D.~Onstad (2014, 01).
\newblock {\em Insect Resistance Management: Biology, Economics, and
  Prediction}.
\newblock Cambridge, Massachusetts, US: Academic Press.

\bibitem[\protect\citeauthoryear{Mulder, Bazeley-White, Dimitrakopoulos,
  Hector, Scherer-Lorenzen, and Schmid}{Mulder et~al.}{2004}]{Mulder2004}
Mulder, C. P.~H., E.~Bazeley-White, P.~G. Dimitrakopoulos, A.~Hector,
  M.~Scherer-Lorenzen, and B.~Schmid (2004).
\newblock Species evenness and productivity in experimental plant communities.
\newblock {\em Oikos\/}~{\em 107\/}(1), 50--63.

\bibitem[\protect\citeauthoryear{Muller, Adriaanse, Belshaw, and
  Godfray}{Muller et~al.}{1999}]{muller1999}
Muller, C., I.~Adriaanse, R.~Belshaw, and H.~Godfray (1999).
\newblock The structure of an aphid--parasitoid community.
\newblock {\em Journal of Animal Ecology\/}~{\em 68\/}(2), 346--370.

\bibitem[\protect\citeauthoryear{Naranjo-Guevara, Santos, Barbosa, Castro, and
  Fernandes}{Naranjo-Guevara et~al.}{2020}]{Guevara2020}
Naranjo-Guevara, N., L.~Santos, N.~Barbosa, A.~Castro, and O.~Fernandes (2020,
  01).
\newblock Long-term mass rearing impacts performance of the egg parasitoid
  \textit{Telenomus remus} (hymenoptera: Platygastridae).
\newblock {\em Journal of Entomological Science\/}~{\em 55}, 69--86.

\bibitem[\protect\citeauthoryear{Noss}{Noss}{1990}]{NOSS1990}
Noss, R.~F. (1990).
\newblock Indicators for monitoring biodiversity: A hierarchical approach.
\newblock {\em Conservation Biology\/}~{\em 4\/}(4), 355--364.

\bibitem[\protect\citeauthoryear{O'Connor, Bojinski, Röösli, and
  Schaepman}{O'Connor et~al.}{2020}]{Oconnor2020}
O'Connor, B., S.~Bojinski, C.~Röösli, and M.~E. Schaepman (2020).
\newblock Monitoring global changes in biodiversity and climate essential as
  ecological crisis intensifies.
\newblock {\em Ecological Informatics\/}~{\em 55}, 1--8.

\bibitem[\protect\citeauthoryear{Ojo and Ola-Adams}{Ojo and
  Ola-Adams}{1996}]{ojo1996}
Ojo, L. and B.~Ola-Adams (1996).
\newblock Measurement of tree diversity in the nigerian rainforest.
\newblock {\em Biodiversity \& Conservation\/}~{\em 5\/}(10), 1253--1270.

\bibitem[\protect\citeauthoryear{Pesenti, Quatto, and Ripamonti}{Pesenti
  et~al.}{2016}]{Pesenti2016}
Pesenti, N., P.~Quatto, and E.~Ripamonti (2016, 10).
\newblock Bootstrap confidence intervals for biodiversity measures based on
  gini index and entropy.
\newblock {\em Quality Quantity\/}~{\em 51}, 847–858.

\bibitem[\protect\citeauthoryear{Pielou}{Pielou}{1966}]{Pielou1966}
Pielou, E.~C. (1966).
\newblock The measurement of diversity in different types of biological
  collections.
\newblock {\em Journal of Theoretical Biology\/}~{\em 13}, 131--144.

\bibitem[\protect\citeauthoryear{Pla}{Pla}{2004}]{Pla2004}
Pla, L. (2004, 03).
\newblock Bootstrap confidence intervals for the shannon biodiversity index: A
  simulation study.
\newblock {\em Journal of Agricultural Biological and Environmental
  Statistics\/}~{\em 9}, 42--56.

\bibitem[\protect\citeauthoryear{Ricklefs}{Ricklefs}{2016}]{Ricklefs2016}
Ricklefs, R.~E. (2016).
\newblock {\em Ecology: The Economy of Nature\/} (7 ed.).
\newblock New York, US: W. H. Freeman.

\bibitem[\protect\citeauthoryear{Scherer, Schaarschmidt, Prescher, and
  Priesnitz}{Scherer et~al.}{2013}]{Scherer2013}
Scherer, R., F.~Schaarschmidt, S.~Prescher, and K.~U. Priesnitz (2013, 03).
\newblock Simultaneous confidence intervals for comparing biodiversity indices
  estimated from overdispersed count data.
\newblock {\em Biometrical journal\/}~{\em 55}, 246–26.

\bibitem[\protect\citeauthoryear{Shenton, Bowman, and Hutcheson}{Shenton
  et~al.}{1969}]{shenton1969}
Shenton, L., K.~Bowman, and K.~Hutcheson (1969).
\newblock Comments on the distribution of indices of diversity.
\newblock In {\em International Symposium of Statistical Ecology}, pp.\
  315--359.

\bibitem[\protect\citeauthoryear{St~Laurent, Mielke, Herbin, Dexter, and
  Kawahara}{St~Laurent et~al.}{2020}]{StLaurent2020}
St~Laurent, R., C.~Mielke, D.~Herbin, K.~Dexter, and A.~Kawahara (2020, 01).
\newblock A new target capture phylogeny elucidates the systematics and
  evolution of wing coupling in sack‐bearer moths.
\newblock {\em Systematic Entomology\/}~{\em 45}, 653--669.

\bibitem[\protect\citeauthoryear{Tsakalakis, Blasius, and Ryabov}{Tsakalakis
  et~al.}{2020}]{Tsakalakis2020}
Tsakalakis, I., B.~Blasius, and A.~Ryabov ({2020}, {JUN}).
\newblock {Resource competition and species coexistence in a two-patch
  metaecosystem model}.
\newblock {\em {THEORETICAL ECOLOGY}\/}~{\em {13}\/}({2}), {209--221}.

\bibitem[\protect\citeauthoryear{Vanclay}{Vanclay}{1989}]{vanclay1989}
Vanclay, J.~K. (1989).
\newblock A growth model for north queensland rainforests.
\newblock {\em Forest Ecology and Management\/}~{\em 27\/}(3-4), 245--271.

\bibitem[\protect\citeauthoryear{von Burg, van Veen, {\'A}lvarez-Alfageme, and
  Romeis}{von Burg et~al.}{2011}]{von2011}
von Burg, S., F.~J. van Veen, F.~{\'A}lvarez-Alfageme, and J.~Romeis (2011).
\newblock Aphid--parasitoid community structure on genetically modified wheat.
\newblock {\em Biology Letters\/}~{\em 7\/}(3), 387--391.

\bibitem[\protect\citeauthoryear{Willett, Filgueiras, Nyrop, and Nault}{Willett
  et~al.}{2020}]{Willett2020}
Willett, D., C.~Filgueiras, J.~Nyrop, and B.~Nault (2020, 04).
\newblock Field monitoring of onion maggot (\textit{Delia antiqua}) fly through
  improved trapping.
\newblock {\em Journal of Applied Entomology\/}~{\em 144\/}(5), 382–387.

\bibitem[\protect\citeauthoryear{Williams, Humphries, Vane-Wright, and
  Gaston}{Williams et~al.}{1996}]{Williams1996}
Williams, P., C.~Humphries, D.~Vane-Wright, and K.~Gaston (1996).
\newblock Value in biodiversity, ecological services and consensus.
\newblock {\em Trends in Ecology \& Evolution\/}~{\em 11\/}(9), 385.

\bibitem[\protect\citeauthoryear{Wilsey and Polley}{Wilsey and
  Polley}{2002}]{wilsey2002}
Wilsey, B.~J. and H.~W. Polley (2002).
\newblock Reductions in grassland species evenness increase dicot seedling
  invasion and spittle bug infestation.
\newblock {\em Ecology Letters\/}~{\em 5\/}(5), 676--684.

\bibitem[\protect\citeauthoryear{Wubs and Bezemer}{Wubs and
  Bezemer}{2018}]{Wubs2018}
Wubs, E. R.~J. and T.~M. Bezemer (2018).
\newblock Plant community evenness responds to spatial plant–soil feedback
  heterogeneity primarily through the diversity of soil conditioning.
\newblock {\em Functional Ecology\/}~{\em 32\/}(2), 509--521.

\bibitem[\protect\citeauthoryear{Zar}{Zar}{2009}]{Zar1999}
Zar, J.~H. (2009).
\newblock {\em Biostatistical analysis\/} ($5^{th}$ ed.).
\newblock Upper Suddle River, New Jersey, USA: Prentice Hall.

\bibitem[\protect\citeauthoryear{Zhang and Zhang}{Zhang and
  Zhang}{2012}]{Zhang2012}
Zhang, Z. and X.~Zhang (2012, 05).
\newblock A normal law for the plug-in estimator of entropy.
\newblock {\em IEEE Transactions on Information Theory - TIT\/}~{\em 58},
  2745--2747.

\end{thebibliography}
\end{document}